\documentclass[prb,twocolumn,noshowpacs,amsmath,twoside]{revtex4}
\usepackage{bm}% bold math
\usepackage{graphicx, amssymb}% Include figure files
\usepackage[T1]{fontenc}

\begin{document}

\title{Glassy dynamics: effective temperatures and intermittencies from a 
two-state model}

\author{M. Naspreda\footnote{Corresponding author: naspreda@ffn.ub.es}}, 
\author{D. Reguera}, 
\author{A. P\'erez-Madrid},
\author{J. M. Rub\'{\i}}

\address{Departament de F\'{\i}sica Fonamental,
Universitat de Barcelona, Diagonal 647, E-08028 Barcelona, Spain}
%\corauth[cor]{Corresponding author.}
%\ead{naspreda@ffn.ub.es}

\date{October 17, 2004}

\begin{abstract}
We show the existence of intermittent dynamics in one of the simplest model of 
a glassy system: the two-state model, which has been used
\cite{Perez&Reguera&Rubi03} to 
explain the origin of the violation of the fluctuation-dissipation theorem. The
dynamics is analyzed through a Langevin equation for the evolution of the state
of the system through its energy landscape. The results obtained concerning the
violation factor and the non-Gaussian nature of the fluctuations are in  good
qualitative agreement with experiments measuring the effective temperature and
the voltage fluctuations in gels and in polymer glasses. The method proposed 
can be useful to study the dynamics of other slow relaxation systems in which 
non-Gaussian fluctuations have been observed.
\end{abstract}

\maketitle

\section{Introduction}

   Complex systems are often distinguished by the existence of a very intricate
free energy landscape consisting of many barriers which the system has to 
overcome to evolve. It is precisely the presence of these barriers the
responsible for the slow relaxation dynamics which manifests in the appearance 
of peculiar phenomena as aging, lack of a fluctuation-dissipation theorem and 
intermittencies caused by the presence of large fluctuations. These 
features, predicted and observed in systems of different nature as glasses, 
granular flows, foams, crumpled materials and in the dynamics of the disordered
systems 
\cite{Rubi&Conrad97,Angel95,Ben-Naim et al98,Sollich97,Matan02,Oliveira02},
have attracted the interest of many researchers during the last years with the 
purpose of describing the main features of slow relaxation dynamics
\cite{Cugliandoloprl}.

   In a previous paper \cite{Perez&Reguera&Rubi03},  we have proposed a 
minimal relaxation model aimed at characterizing the dynamics of a system 
relaxing in two very different time scales, which are related to inter-well 
and intra-well relaxation processes. Two scenarios were analyzed for this 
purpose. In the first of them, the system may explore the whole reaction 
coordinate space undergoing a diffusion process described by a Fokker-Planck 
equation \cite{Mohanty94}, which accounts for the intra-well
and inter-well relaxations. In  the other,  obtained from the first one by 
eliminating the fast variable, the system undergoes an activated process. In 
spite of its simplicity the model shows some of the peculiar features of the 
dynamics of slow relaxation systems and proposes an explanation of  why and how
the fluctuation-dissipation is violated. It was found that the violation factor
or effective temperature depends on the observable and on the initial 
populations in the wells. This result shows that the effective temperature 
does not univocally characterize the thermal state of a glassy system 
undergoing activated dynamics \cite{Vilar01}.

    Our purpose in this paper is to use that model to explain the presence of 
intermittencies in the dynamics of a system in a glassy phase and the
non-Gaussian nature of the probability distribution function, which have
recently been observed in measurements of the dielectric properties of gels and
polymer glasses \cite{Ciliberto1,Ciliberto2}, and in some theoretical studies
of spin-glass models \cite{Felix04,Sibani04}.
 
    The paper is organized as follows. In Section 2, we analyze some of the 
main traits of glassy dynamics from a two-state model. Section 3 is devoted to 
present the results concerning the intermittent behavior and the non-Gaussian
nature of the fluctuations. Some conclusions and perspectives are presented in 
the final section.

\section{Glassy dynamics from a two-state model}

    In a two-state model, the minimal relaxation model for glassy systems
\cite{Huse86,Langer88,Langer90}, one assumes that the process consists of two 
main steps: a slow relaxation, in which the coordinate characterizing the state
of the system jumps from a potential well to the next one, and a fast
equilibration process in the well. It has been shown
\cite{Perez&Reguera&Rubi03} that the dynamics of the system can be analyzed in 
terms of a Fokker-Planck equation describing a diffusion process through the 
free energy landscape $\Phi(\gamma)$ \cite{Ignacio97},

\begin{equation}\label{F-P}
\frac{\partial \rho(\gamma,t)}{\partial t} =
\frac{\partial}{\partial\gamma}D\left[\frac{\partial\rho(\gamma,t)}{\partial
\gamma} + \frac{\rho(\gamma,t)}{k_BT} \frac{\partial\Phi(\gamma)}{\partial
\gamma}\right].
\end{equation}

In the simplest case, $\Phi(\gamma)$ would be just a bistable potential. In the
previous equation, $\rho(\gamma,t)$ is the probability distribution function 
which depends on the order parameter or reaction coordinate $\gamma$, $D$ is 
the diffusion coefficient, $T$ is the temperature of the bath and $k_B$ the
Boltzmann constant. When the height of the 
barrier separating the two minima of the potential is large enough compared to 
thermal energy the systems achieves a state of quasi-equilibrium in each well. 
The evolution of the system then proceeds by jumps from one well to the other 
undergoing an activated process. The dynamics corresponding to this situation 
can be obtained by eliminating the fast degrees of freedom in such a way that 
it can be characterized simply by the populations at each well.  The 
Fokker-Planck equation then reduces to the following kinetic equations 
governing the population dynamics at both wells 
\cite{Perez&Reguera&Rubi03,Schmid01}

\begin{equation}\label{master equation}
\frac{dn_1}{dt} = -\frac{dn_2}{dt} = - j(t) - j^r(t).
\end{equation}
where the current $j(t)$ is defined as:

\begin{equation}
j(t) = j_\rightarrow - j_\leftarrow = 
\Big(k_\rightarrow n_1 - k_\leftarrow n_2\Big)
\end{equation}

Here $n_1(t)$ and $n_2(t)$ are the populations at each well and 
$k_{\rightarrow,\leftarrow}$ are the forward and backward reaction rates given 
by

\begin{equation}
k_{\rightarrow, \leftarrow} = \frac{D\sqrt{\Phi''(\gamma_{1,2})
|\Phi''(\gamma_0)|}}{2\pi k_BT}\exp\left[\frac{\Phi(\gamma_{1,2})-
\Phi(\gamma_0)}{k_BT}\right],
\end{equation}
where $\gamma_1$ and $\gamma_2$ denote the position of the two minima and 
$\gamma_0$ the position of the top at the barrier. The noise term $j^r(t)$ 
results from the coarsening of the corresponding random term $J^r(\gamma,t)$
in the Langevin equation related to the Fokker-Planck equation (\ref{F-P}), 
\cite{Perez&Reguera&Rubi03}. Whereas the latter 
satisfies the fluctuation-dissipation theorem inherent to the diffusion 
process along the reaction coordinate  formulated as \cite{Landau}

\begin{equation}
\langle J^r(\gamma,t)J^r(\gamma',t') \rangle = 
2D\langle \rho(\gamma,t)\rangle
\delta(\gamma-\gamma')\delta(t-t')
\end{equation}

the former does not obey a similar expression. Its correlation is given by
\cite{Ignacio97}:

\begin{equation}
\langle j^r(t)j^r(t')\rangle = (k_\rightarrow \langle n_1\rangle +
k_\leftarrow \langle n_2\rangle) \delta(t-t') \neq 2k_\rightarrow n_1^{eq}
\delta(t-t')
\end{equation}
where the last term is the value of the correlation at equilibrium, and 
$n_1^{eq}$ is the equilibrium population in the first well. 

    This result clearly indicates that the violation of the fluctuation-
dissipation theorem is precisely due to the coarsening of the description. 
When one eliminates the fast variables reducing the dynamics to that of an 
activated process the system cannot progressively pass through local 
equilibrium states from one well to the other but proceeds by jumps and 
consequently remains outside equilibrium. This explains why the fluctuation-
dissipation theorem, a result strictly valid when the fluctuations take place 
around an equilibrium state \cite{Callen&Greene}, is not 
fulfilled.

   From the model proposed we can analyze the non-equilibrium response of the 
system. Let us assume an external perturbation $-\varepsilon(t)O(t)$ that is
plugged in at the waiting time $t_w$, defined as the time elapsed after
quenching ($O(t)$ is the observable considered and $\varepsilon(t) =
\varepsilon_0 \theta(t-t_w)$ is a generelized force that is coupled to the 
observable). The response is then given by

\begin{equation}
R(t,t_w) = \left.\frac{\partial \langle \delta O(t)\rangle}{\partial\varepsilon
(t_w)}\right|_{\varepsilon_0\rightarrow 0},
\end{equation}
where the average of the observable is defined as

\begin{equation}
\langle \delta O(t)\rangle = \int O(\gamma)\delta \rho(\gamma,t)d\gamma =
\left( O_1 - O_2\right)\delta n_1,
\end{equation}
whereas the values $O_1$ and $O_2$ are the values of the observable at the
minima 1 and 2, respectively. On the other hand, the correlation is

\begin{equation}
C_O(t,t_w) = e^{-(t-t_w)/\tau} (O_1 - O_2) [j_\rightarrow (O_1 - O_0) - 
j_\leftarrow (O_2 - O_0)].
\end{equation} 

     Both quantities satisfy the relation

\begin{equation}\label{FDT}
R_O(t,t_w) = \frac{1}{k_B T_{eff}^O}\frac{\partial}{\partial t_w}C_O(t,t_w),
\end{equation}
in which the quantity $T_{eff}^O$ plays the role of an ``effective'' 
temperature given by

\begin{equation}\label{effective temperature}
T_{eff}^O = \frac{T}{Ae^{-t_w/\tau} + (1-e^{-t_w/\tau})}.
\end{equation}
 
Here $\tau^{-1}=(k_\rightarrow + k_\leftarrow)$ is the relaxation time of
the process and the parameter $A$ has the form

\begin{figure}[t]
\begin{center}
\includegraphics[height = 2.5in, width = 3.5in, angle = 0, origin = c]{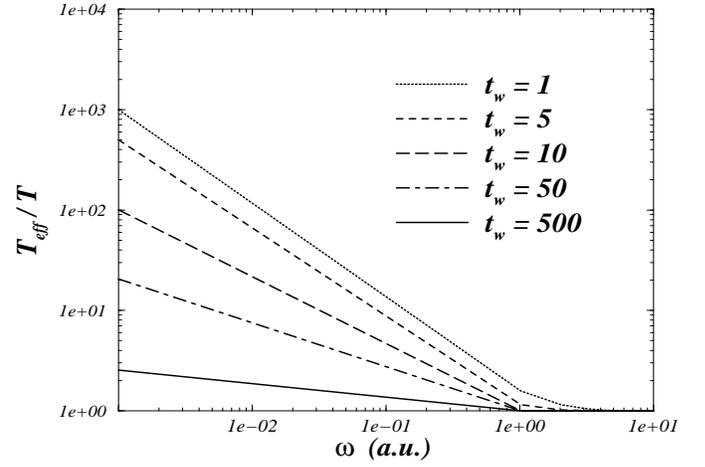}
\caption{The effective temperature plotted as a function of the 
frequency (in arbitrary units) for different values of the waiting time.}
\end{center}
\label{effective temperature2}
\end{figure}

\begin{equation}\label{A parameter}
A = \frac{k_\rightarrow \langle n_1(0)\rangle (O_1 - O_0) - 
k_\leftarrow \langle n_2(0) (O_2 - O_0)}{k_\rightarrow n_1^{eq}(O_1 - O_2)}
\end{equation}

Since the equilibrium state is achieved for $t_w\rightarrow \infty$, one 
can easily verify that the effective temperature coincides with that of the 
bath and the relation (\ref{FDT}) becomes the fluctuation-dissipation 
theorem. 

    An important consequence of our previous analysis is  that the effective 
temperature is not a robust quantity since it depends on the observable 
considered as well as on the initial populations at the wells. We then conclude
that in the case of an activated process that quantity is a parameter 
measuring the distance of the system to the equilibrium state but it is not
universal and consequently can hardly be considered as a 
thermodynamic temperature \cite{Vilar01}.

     Our expression Eq.(\ref{effective temperature}) can be used to reproduce 
the dependence of the violation factor on the frequency observed in experiments
\cite{Ciliberto1,Ciliberto2}. If we assume that the whole relaxation process
takes place through consecutive activated processes for which Eq.(\ref{master 
equation}) applies, we can infer a global behavior of that quantity by 
identifying the inverse of the relaxation time with a frequency. By considering
the case $A=0$, in which the violation factor does not depend on the observable
, we then conclude that the effective temperature at low frequencies behaves
as: $T_{eff} \sim \omega^{-1}$. In the experiments one finds that the effective
temperature decreases when increasing the frequency following a power law 
whose exponent is in between $-1.1$ and $-1.2$, \cite{Ciliberto2}. As in the 
experiments, it is found that the effective temperature tends to the bath 
temperature for very large waiting times (see Fig. 1).

\section{Intermittent dynamics}

    We will analyze in this section the intermittent behavior of the 
fluctuations taking place when the system is quenched below the glass 
transition. This type of behavior has been reported in experiments measuring 
the fluctuation spectrum and the response of a Laponite solution and of a 
polymer glass \cite{Ciliberto1,Ciliberto2}. To this purpose, we will 
particularize the model discussed in Sec. 2 to the case of the quartic
potential plotted in Fig. 2. The dynamics of this model is
governed by the Fokker-Planck equation (\ref{F-P}) or through the
equivalent Langevin equation

\begin{figure}[b]
\begin{center}
\includegraphics[height = 2.5in, width = 3.5in, angle = 0, origin = c]{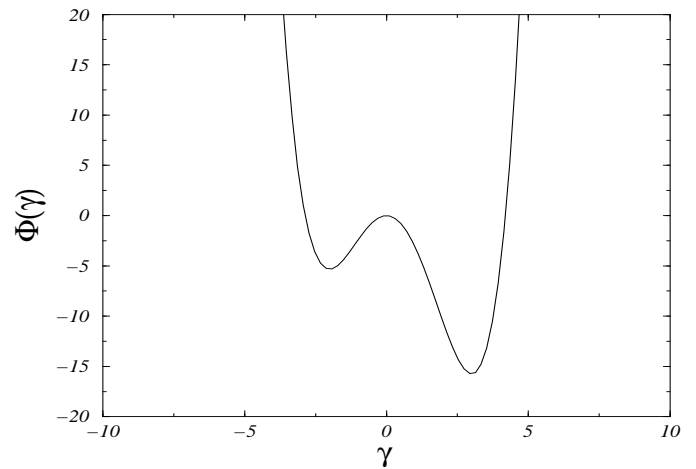}
\caption{Quartic potential used in our model.}
\end{center}
\label{potential}
\end{figure}

\begin{equation}\label{SDE}
\frac{d\gamma}{dt} = -\gamma (\gamma - \gamma_1) (\gamma - \gamma_2) 
+ J^r.
\end{equation}
where $J^r$ has been considered a Gaussian white noise process having,
as a first approximation, a constant amplitude: $D\langle\rho(\gamma,t)\rangle
\approx \alpha$. We will see that this approximation is enough to explain the 
experimental results. 

\begin{figure}[t]
\begin{center}
\includegraphics[height = 2.5in, width = 3.5in, angle = 0, origin = c]{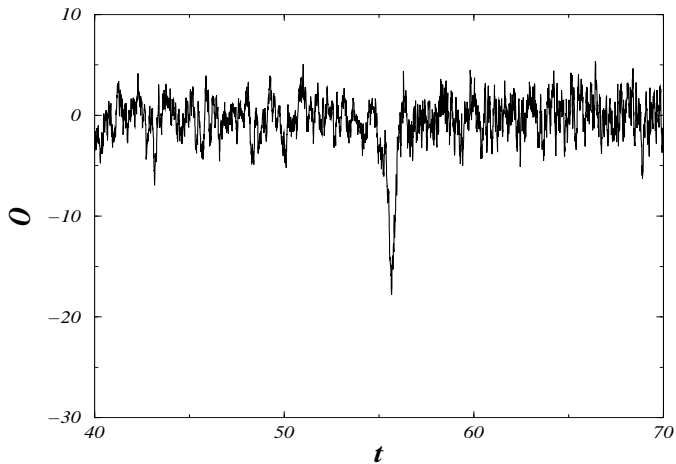}
\caption{The evolution of a single realization of the observable $O(\gamma)$,
corresponding to $\alpha = 1.5$. The single step-time has been taken as 0.01. 
We have considered 10000 time intervals.}
\end{center}
\label{observable}
\end{figure}

\begin{figure}[b]
\begin{center}
\includegraphics[height = 2.5in, width = 3.5in, angle = 0, origin = c]{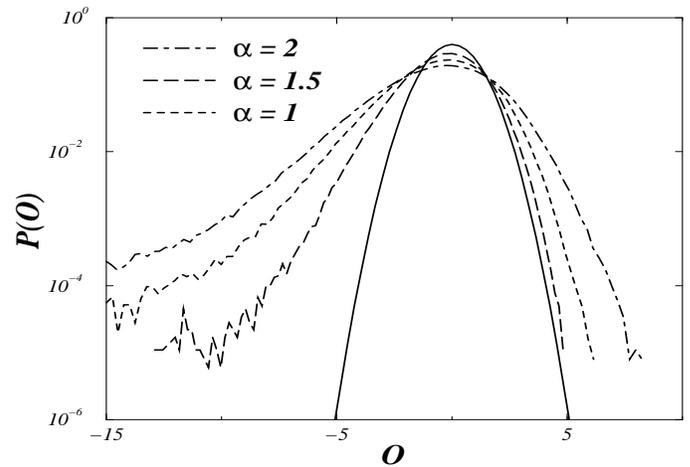}
\caption{Probability distribution function for different values of $\alpha$ as
a function of the values of the observable, obtained by performing
simulation of Eq.(\ref{SDE}). The solid line is a
Gaussian curve plotted just for comparison.}
\end{center}
\label{pdf}
\end{figure}

   By numerical simulation of the Langevin equation (\ref{SDE}), we have 
computed the value of the following observable

\begin{equation}
O(\gamma) = \left\{\begin{array}{rl} -a_1 (\gamma - \gamma_1) &\qquad 
\textrm{if    } \gamma < 1 \\ a_2 (\gamma - \gamma_2)& \qquad 
\textrm{if   } \gamma > 
1\end{array}\right.,
\end{equation}
which has a piece-wise dependence on $\gamma$, having a zero value at
$\gamma_1$ and $\gamma_2$. In our case we have taken $\gamma_1 = -2$ and
$\gamma_2 = 3$ and the parameters $a_1$ and $a_2$ have been chosen arbitrarily
as $a_1 = 6$ and $a_2 = 9$, ensuring the continuity of the observable. The
location of the maximum of the observable has been placed at $\gamma = 1$, to
show in a clearer way the occurrence of a transition between the two wells.
When the state is at the top of the barrier the probability to go
back to the minimum $\gamma_1$ is 50\%.

    The value of $\alpha$ should be consistent with the non-stationary nature
of the process and has to be intrinsically related to the waiting time. 
Therefore, we have use different values of this parameter to mimic different
values of the waiting time. 

   In Fig. \ref{observable} we have represented the value of the observable 
$O(\gamma)$ corresponding to a single trajectory obtained from the simulations 
of Eq.(\ref{SDE}), with $\alpha = 1.5$. We observe the presence of an 
intermittent event, which is the signature of a jump from one well to another.
Therefore, the presence of intermittencies is a consequence of the activated
nature of the relaxation process.

   Averaging over many trajectories we have obtained the probability
distribution function for the observable, which has been plotted in Fig. 4,
for different values of $\alpha$. The non-Gaussian form of the curves is,
in our model, a consequence of the non-parabolic form of the potential.
Intermittencies add more weight to the tails of the distribution, thus
stressing its non-Gaussian nature.

\section{Conclusions}
    In this paper we have used the activation over a  barrier to model the 
evolution of a system when is  quenched below the glass transition. Using as a
 model a simple quartic potential, we have proved the existence of an 
intermittent dynamics in the aging process caused by the presence of large 
fluctuations. Similar behavior has been observed in recent experiments. We 
have shown that the probability distribution function corresponding to those 
events deviates from its Gaussian form observed for systems close to 
equilibrium and exhibits exponential tails as those encountered experimentally.

    The model proposed and its possible generalizations could, with the help of
the stochastic processes theory \cite{Hanggi90,Talkner04},
constitute a useful tool to characterize the dynamics of systems far from 
equilibrium for which a common phenomenology in the behavior of the 
fluctuations is being found.

\section*{Acknowledgments}
This work was partially supported by the DGICYT and FEDER under 
Grant No.\ BFM2002-01267.
D. R. acknowledges support by the Ministerio de Ciencia y
Tecnolog\'{\i}a of Spain through the ``Ram\'on y Cajal'' program.

\end{document}